\documentclass[10pt,letterpaper]{article}
\usepackage[top=0.85in,left=2.75in,footskip=0.75in]{geometry}

\usepackage{amsmath,amssymb}

\usepackage{changepage}

\usepackage[utf8x]{inputenc}
\usepackage{rotating}
\usepackage{textcomp,marvosym}

\usepackage{cite}

\usepackage{nameref}
\usepackage{hyperref}
\usepackage[right]{lineno}

\usepackage{microtype}
\DisableLigatures[f]{encoding = *, family = * }

\usepackage[table]{xcolor}

\usepackage{array}

\newcolumntype{+}{!{\vrule width 2pt}}

\newlength\savedwidth

\raggedright
\usepackage[aboveskip=1pt,labelfont=bf,labelsep=period,justification=raggedright,singlelinecheck=off]{caption}

\makeatletter
\renewcommand{\@biblabel}[1]{\quad#1.}
\makeatother

\date{}

\usepackage{lastpage,fancyhdr,graphicx}
\usepackage{epstopdf}
\fancyheadoffset[L]{2.25in}
\fancyfootoffset[L]{2.25in}
\begin{document}
\vspace*{0.2in}

\begin{flushleft}
{\Large
% \textbf\newline{Distance Matters: Quantifying the Impact of Scholarly Papers Based on Geographical Distance} % Please use "sentence case" for title and headings (capitalize only the first word in a title (or heading), the first word in a subtitle (or subheading), and any proper nouns).
\textbf\newline{Quantifying the Impact of Scholarly Papers Based on Higher-Order Weighted Citations} % Please use "sentence case" for title and headings (capitalize only the first word in a title (or heading), the first word in a subtitle (or subheading), and any proper nouns).
}
\newline
\\
Xiaomei Bai\textsuperscript{1,2},
Fuli Zhang\textsuperscript{3},
Jie Hou\textsuperscript{2},
Ivan Lee\textsuperscript{4,2*},
Xiangjie Kong\textsuperscript{2},
Amr Tolba\textsuperscript{5,6},
Feng Xia\textsuperscript{2}
\\
\bigskip
\bf{1} Computing Center, Anshan Normal University, Anshan 114007, China
\\
\bf{2} Key Laboratory for Ubiquitous Network and Service Software of Liaoning Province, School of Software, Dalian University of Technology, Dalian 116620, China
\\
\bf{3} Library, Anshan Normal University, Anshan 114007, China
\\
\bf{4} School of ITMS, University of South Australia, Mawson Lakes, SA 5095, Australia
\\
\bf{5} Computer Science Department, Community College, King Saud University, Riyadh 11437, Saudi Arabia
\\
\bf{6} Mathematics Department, Faculty of Science, Menoufia University, Shebin-El-Kom 32511, Egypt
%\\
\bigskip

% Insert additional author notes using the symbols described below. Insert symbol callouts after author names as necessary.
%
% Remove or comment out the author notes below if they aren't used.
%
% Primary Equal Contribution Note
%\Yinyang These authors contributed equally to this work.
%
%% Additional Equal Contribution Note
%% Also use this double-dagger symbol for special authorship notes, such as senior authorship.
%\ddag These authors also contributed equally to this work.
%
%% Current address notes
%\textcurrency Current Address: Dept/Program/Center, Institution Name, City, State, Country % change symbol to "\textcurrency a" if more than one current address note
%% \textcurrency b Insert second current address
%% \textcurrency c Insert third current address
%
%% Deceased author note

% Group/Consortium Author Note
%\textpilcrow Membership list can be found in the Acknowledgments section.

% Use the asterisk to denote corresponding authorship and provide email address in note below.
* ivan.lee@unisa.edu.au

\end{flushleft}

\section*{Abstract}
Quantifying the impact of a scholarly paper is of great significance, yet the effect of geographical distance of cited papers has not been explored. In this paper, we examine 30,596 papers published in Physical Review C, and identify the relationship between citations and geographical distances between author affiliations. Subsequently, a relative citation weight is applied to assess the impact of a scholarly paper. A higher-order weighted quantum PageRank algorithm is also developed to address the behavior of multiple step citation flow. Capturing the citation dynamics with higher-order dependencies reveals the actual impact of papers, including necessary self-citations that are sometimes excluded in prior studies. Quantum PageRank is utilized in this paper to help differentiating nodes whose PageRank values are identical.

\section*{Introduction}
With the rapidly growth of scholarly big data~\cite{xia2017big}, there's a crucial need to quantify the impact of scholarly papers, to assess the performance of individual scholars, institutions, even for countries~\cite{aguinis2012scholarly}. Currently, the impact of scholarly paper is mainly divided into two categories: unstructured metrics and structured metrics~\cite{Bai2017An}. Unstructured metrics evaluate the impact of scholarly paper from a statistical point of view. Citations~\cite{evans2009open}~\cite{gargouri2010self} are the most representative unstructured metrics, with examples such as the H-index~\cite{hirsch2005index}, the g-index~\cite{egghe2006theory}, and the impact factor (IF)~\cite{garfield2006history}.
As an alternative measure of scientific impact, Xia et al.\cite{xia2015bibliographic} have investigated scholarly impact reflected on social media, and explore the correlation between citations and messages/tweets on Facebook and Tweeter. The structured metrics mainly consider the importance of scholarly entities in scholarly network, such as citation network, co-authors network, author-paper network, etc. PageRank~\cite{brin1998the}, a seminal example of structured metrics, has attracted growing attentions in scholarly impact evaluation. Sayyadi et al.~\cite{sayyadi2009futurerank} have estimated future prestige scores of scholarly papers via the following three features: citations, publication date, and authorship. Wang et al.~\cite{wang2013ranking} have quantified the impact of scholarly papers by applying PageRank and HITS~\cite{kleinberg1999authoritative} on citation network, author-paper network, and journal-paper network. In the unweighted structured metrics, all citations are treated with equal importance. An alternative approach is to evaluate the impact of scholarly papers by time-aware weighted citation network~\cite{wang2014future}. In another study, Shah et al.~\cite{shah2015s} has proposed the S-index metric to model the influence prorogation by a weighted paper-paper citation networks. This paper applies a hierarchical model between the citing paper and the cited paper, thus the impact of a scholarly paper decayed rapidly over different hierarchical levels.

One potential problem for unstructured and structured metrics is that the impact of individual papers can be manipulated. For instance, aggressive self-citations or induced-citations may lead to an inflated impact. Bai et al.~\cite{bai2016identifying} has evaluated the impact of scholarly papers using a weighted citation network, in which Conflict of Interest citation relationships are identified and the citation strengths are weakened. Another potential problem with structured metrics is that little is known how actual geographic distance influences the impact of scholarly paper, and how higher-order dependencies in citation networks react to the impact of scholarly paper. Liben-Nowell et al.~\cite{Liben2005Geographic} investigated the relationship between geographical distance and friendship in the LiveJournal network, indicating that geographical proximity can indeed increase the probability of friendship. This proved that social network attributes and geographical distance is related, which is an important aspect of the theory of small world. A previous research found a strong linear relationship between institutions and distance~\cite{B2006Mapping}. Schubert et al.~\cite{Schubert2006Cross} revealed that geopolitical location, cultural relations and language are important factors in shaping preference of cross-citation. Wu~\cite{Wu2013Geographical} investigated citing distances, citation patterns and spatial diversity to explore geographical knowledge diffusion. Albarran et al.~\cite{Albarr2014Differences} found economic, political, sociological and intellectual factors were influencing the shaping of their citation distributions and the research performance of countries. A geographic analysis of citation flows between cities is helpful to uncover how new scientific paradigms spread, and understand how quickly a new research gets recognized by academic circle in different geographical areas ~\cite{Pan2011The}.
Bai et al.~\cite{bai2017evaluating} explored the relationships between citations and the actual geographic location of institutions for evaluating the impact of scholarly papers. Based on the previous work, we further explore the relationships between them, and construct a relative weight to represent the importance of citation.

The concept of higher-order dependencies has been introduced by Xu et al.~\cite{Xu2015Representing} to ensure the correctness of network analysis. The higher-order dependencies mean that, when movements are simulated on the network, the next movement depends on several previous steps. The higher-order dependencies are widely applied to model various applications, including Web browsing behaviors~\cite{Deshpande2004Selective}, vehicle and human movements~\cite{Rosvall2014Memory}, stock market~\cite{janacek2010time}, etc. Bohlin et al.~\cite{Bohlin2015Robustness} have modelled citation flow between journals, and remembered their previous steps, corresponding to the zero-, first-, and second-order Markov models. Previous researchers evaluate the impact of papers based on the original citation network, ignoring the influence of multiple step citation flow on the impact of papers. In this paper, we construct a higher-order citation network, and apply the hierarchical citation structure to quantify the impact of scholarly papers.

Once a citation network is constructed, evaluation methods such as PageRank or HITS can be applied. Although PageRank was introduced to rank Web pages, the algorithm has been deployed in many applications such as finding important nodes in networks~\cite{Gleich2014PageRank}, measuring impacts of scholarly papers~\cite{chen2007finding}, evaluating impacts of scholars~\cite{zhang2016rising} or journals~\cite{Chen2011An}, as well as various applications in social networks~\cite{Dou2016Beyond} and graph analysis~\cite{Kalavri2016The}. A personalized PageRank was developed to find the vertices in a graph~\cite{xie2015edge}. A multilinear PageRank modified the PageRank to a higher-order Markov chain, and studied a computationally tractable approximation to the higher-order PageRank vector~\cite{Gleich2014Multilinear}. The multilinear PageRank modelled stochastic processes depending on the previous steps. But, it is well known that using PageRank to evaluate scholarly impact brings a problem of the evaluation results depending on parameter $\alpha$. When the $\alpha$ value is different, the evaluation results will be changed accordingly.

To address the limitation of PageRank, Paparo et al.~\cite{Paparo2014Quantum} have proposed the quantum PageRank algorithm to unambiguously identify the underlying topology of networks. The quantum PageRank algorithm clearly highlights the structure of secondary hubs in scale-free networks. It recognizes the hierarchical structure in scale-free networks, amplifying the difference of important degree of nodes. The algorithm mainly consists of the following parts: (1) The input state of the algorithm is constructed based on the transition matrix of PageRank. (2) Construct the unitary matrix and transfer matrix to generate the total transformation matrix. (3) In order to obtain the probability of particle appearing in each node, square of total transformation matrix is used to update the initial state. (4) Calculate $m$ times average value for each node of a given network, namely, the quantum PageRank value.

This paper analyzes temporal and geographical attributes of publications and citations, addresses the limitation of conventional techniques in quantifying the impact of scholarly papers, and the main contributions of this paper are summarized as follows:
(1) Identifying the relationship between citations and geographic locations of affiliations.
(2) Introducing a relative citation weight based on geographical distance between institutions to better quantify the impact of scholarly papers.
(3) Exploring higher-order dependencies in citation networks.
(4) Developing the higher-order weighted quantum PageRank algorithm to rank the impact of scholarly papers.

\section*{Methods}

\subsection*{Citation between institutions}
\begin{figure}[!h]
\centering
\includegraphics[width=1.0\textwidth]{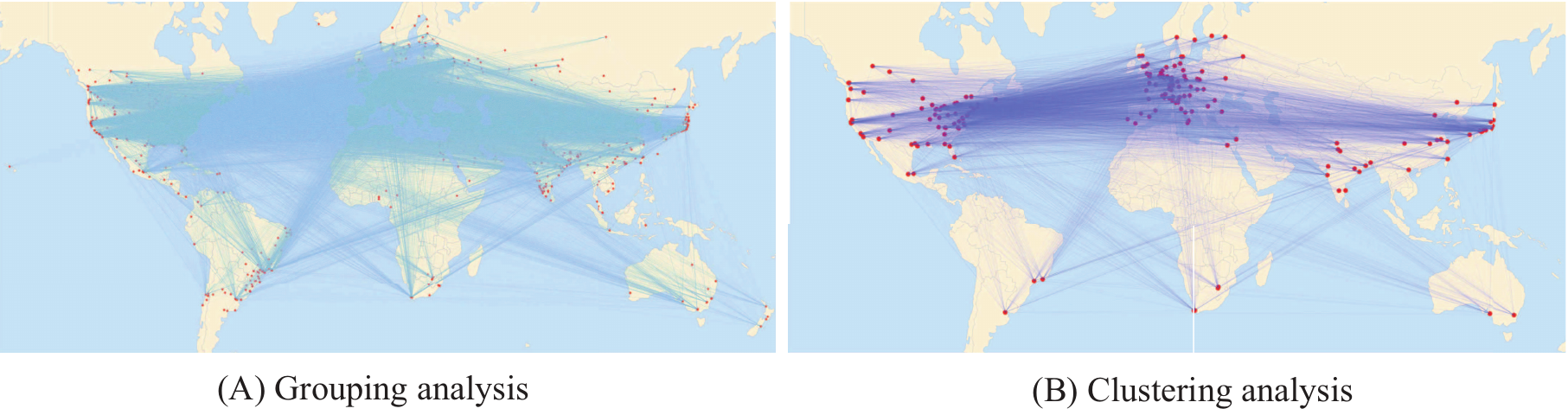}
\caption{Visualizing citations between institutions.}
\label{Fig.1}
\end{figure}

Fig~\ref{Fig.1} shows the citation relationship between institutions using two statistical analysis techniques: grouping analysis~\cite{silbersweig1995functional,venkatramanan2015comprehensive} and clustering analysis~\cite{diday1980clustering}. Red dots represent institutions, and the links between institutions represent citations. Fig~\ref{Fig.1}A shows that the citation relationship between different institutions by grouping analysis.
%The number of institutions is about 1500 by grouping analysis.
The number of institutions is about 200 by clustering analysis. As Fig~\ref{Fig.1} shows, the institutions between six continents cite each other. In particular, the citation between North America and Europe is more frequent compared to between other continents in the field of physics. Fig~\ref{Fig.1}B can more clearly show the frequency of citation.

\subsection*{Relative citation weight}
\textbf{geographical distance}: Let $I$ represent a set of institutions, $I=\{\ I_{1}, I_{2}, \cdots, I_{a}, \cdots, I_{b} \cdots\}$, and $D$ represent the geographic distance between two institutions $I_{a}$ and $I_{b}$. By approximating the geographic distance using the spherical model, $D$ can be formulated as:

\begin{eqnarray}\label{eq:e1}
  D_{I_{a},I_{b}}=2R \cdot \left| arcsin\sqrt{sin^2{\left( \frac{ \left| \Delta\theta \right| }{2} \right)}+cos(\theta_{a})\cdot cos(\theta_{b})\cdot sin^2{\left(\frac{\left|\Delta\phi\right|}{2}\right)}}  \right| ,
\end{eqnarray}
where $R$ is the radius of the earth, $\theta_{a}$ and $\theta_{b}$ are the latitudes of $I_{a}$ and $I_{b}$, $\phi_{a}$ and $\phi_{b}$ are the longitudes of $I_{a}$ and $I_{b}$. $\Delta\theta$ is the differences of latitudes between $I_{a}$ and $I_{b}$, $\Delta\theta=\theta_{a}-\theta_{b}$. $\Delta\phi$ is the difference of longitudes between $I_{a}$ and $I_{b}$, and $\Delta\phi=\phi_{a}-\phi_{b}$.

While physical distance increases communication barrier for physical interactions between the author and citing researcher, it is expected that citation counts decline over the geographical distance that separates the researchers. (Further discussions can be found in the Discussion Section.) The decline pattern is modelled with an exponential decay, according to the following equation:
\begin{eqnarray}\label{eq:e3}
  y = y_{0}+A_{1}e^{\frac{-x}{t_{1}}},
\end{eqnarray}
where $y$ represents the citation count, $y_0$ is a constant representing an offset of the citation count, $x$ is the physical distance separating the researchers, whereas $t_{1}$ represents a scaling factor. $A_{1}$ is the default number of citation less the offset when the author and the citing researcher co-locate at the same physical location. Experimental results of the citation pattern are presented in the Results Section.

Upon identifying the citation pattern, we construct a relative citation weight to quantify the impact of scholarly papers. We consider the citation network at the institution-level, in which each institution has its actual latitude and longitude. Institutions are identified with nodes, and an edge exists between two institutions if they have citation relationships. In the citation network, the relative citation weight between two institutions, $W_{I_{a},I_{b}}$, is defined as:
%\begin{equation}
% W_{I_{a},I_{b}}=\frac{D_{I_{a},I_{a}}}{\frac{1}{2}\sum_{m=1}^{S}\sum_{n=1}^{S}D_{I_{m},I_{n}}}
%\end{equation}
\begin{eqnarray}
  W_{I_{a},I_{b}}=\frac{D_{I_{a},I_{b}}}{\underset{m,n\in G}\max D_{I_{m},I_{n}}},
\end{eqnarray}
where $G$ represent the set of all institutions, and $I_m$ and $I_n$ denotes any two individual institutions and $m \neq n$. $D_{I_{m},I_{n}}$ represents the geographic distance of two different institutions. $\max D_{I_{m},I_{n}}$ indicates the maximum geographic distance between institutions.

\subsection*{Higher-order weighted quantum PageRank}
In this section, we introduce the proposed higher-order quantum PageRank algorithm.
%The idea is mainly inspired by the two papers.
Firstly, we construct higher-order dependencies in citation network. The specific process is as follows: (1) We use the random walk method to find the citation chain from the original citation network to identify the higher-order dependence of the citation relationships among the papers. (2) We traverse all citation chains and add up the number of occurrences of each order citation of all nodes in the chain. Citation chains can navigate backwards and forwards to build up a picture of the intellectual base about a topic~\cite{Cawkell1998Checking}. (3) In the case of different orders, we calculate the probability of each node citing other articles separately. (4) In the different orders, we compare the probability of occurrence in the same citation relationship. If the probability change is large in different orders, the original citation relationship is replaced by the high-order relationship. At the same time, the node representing the higher-order relationship replace the original node. (5) Rewire the higher-order citation edges. It is necessary because a higher-order node replacing the original node can result in the loss of previous steps. (6) Establish the probability transfer matrix $G$ according to all the generated citation relationships.

Given a directed graph with $M$ nodes, $i|k$ indicates the $k$th order of node $i$. $N_{i\rightarrow j}$ indicates the number of occurrences that node $i$ cites node $j$. The probability of node $i$ transferring to its neighboring nodes is defined as:
\begin{eqnarray}
P_{i| k\rightarrow j}=\frac{N_{i| k\rightarrow j}}{\sum_{t=1}^{M}N_{i| k\rightarrow t}},
\end{eqnarray}
where $k\in [2,order]$ with $order$ shows the highest order, and $t$ ranges from $1$ to $M$.

In order to calculate the probability, the K-L divergence value $D(P_i)$ needs to be obtained:
\begin{eqnarray}
D(P_i)=\sum_{j=1}^{M}P_{i|k\rightarrow j}\log \frac{P_{i|k\rightarrow j}}{P_{i\rightarrow j}}.
\end{eqnarray}
If K-L divergence value $D(P_i)$ of node $i$ is bigger than $\frac{k}{log\sum_{t=1}^{M}N_{i|k\rightarrow t}}$, using $i|k$ replaces the previous node $i$, and node $i$ will obtain an updated transition probability.

Secondly, we calculate the transfer matrix $G$ according to the directed graph with $N$ nodes. Subsequently, we need to construct the initial state, namely, input state. The detail is as follows: (1) $|i\rangle|j\rangle$ represents the direct edge that the node $i$ points the node $j$. $G_{k,i}$ indicates the probability of node $i$ to node $k$, where $i, k \in \left [0,N-1 \right ]$. (2) we calculate the superposition of all nodes according to the following formula:
\begin{eqnarray}
|\psi_{j}\rangle:=|j\rangle\otimes\sum_{k=0}^{N-1}\sqrt{G_{kj}}|k\rangle ,
\end{eqnarray}
where $|\psi_{j}\rangle$ indicates a superposition of the vectors, which represents outgoing edges from node $j$.

The stochastic pattern of the vectors $|\psi_{j}\rangle$ for $j=0,1,2,\ldots,N-1$ are normalized in matrix $G$. These vectors form an $N$-dimensional orthonormal set of vectors, and they are used as the initial state of quantum walk.

Then, we need to construct the unitary matrix $\pi$ and the transfer matrix $S$ to obtain the general transform matrix. The unitary matrix $\pi$ is
\begin{eqnarray}
  \pi=2\sum_{j=0}^{N-1}|\psi_{j}\rangle\langle\psi_{j}|-E.
\end{eqnarray}
The transfer matrix $S$ is used to move a quantum particle from node $j$ to node $k$:
\begin{eqnarray}
  S=\sum_{j,k =0}^{N-1}|jk\rangle\langle kj|.
\end{eqnarray}

The general transform matrix is defined as
\begin{eqnarray}
  U=\pi S.
\end{eqnarray}

As the directions of the edges of the graph need to be swapped for an even number of times, we use $U^{2}$ to update the initial state $|\psi_{0}\rangle$ each time. Then we calculate the probability that the particle appears on node $i$. The probability that the particles will appear at node $i$ after $m$ times of walking, $P_{i,m}$, can be obtained using the following formula:
\begin{eqnarray}
P_{i,m}=\langle\psi _{i}|U^{2m\dag}\cdot U^{2m}|\psi _{i}\rangle,
\end{eqnarray}
where  $U^{2m}$ indicates $U^{2}$ iteration $m$ times, $U^{2m\dag}$ is the transpose of $U^{2m}$.

Finally, in order to guarantee a probabilistic interpretation of high order quantum PageRank, we conduct the following process
\begin{eqnarray}
\sum_{i=0}^{N-1}P_{i,m}=\sum_{i=0}^{N-1}\langle\psi _{i}|U^{2m\dag}\cdot U^{2m}|\psi _{i}\rangle = 1, \forall m.
\end{eqnarray}
$P_{i,m}$ can be interpreted as the relative importance degree of node $i$, and it can be found by calculating the probability of a quantum walker on node $i$.
Thus, the impact score of each scholarly paper can be calculated from the $P_{i,m}$ value, as shown in Eq~\eqref{eq:12}.

\subsection*{Definition of a scholarly paper impact}
Based on the observation that citations are inversely related to the geographical distance following an exponential distribution, the impact of each scholarly paper is defined as its average higher-order weighted quantum PageRank value:
\begin{eqnarray}
  S( P_{i})=\langle P_{i,m}\rangle:=\frac{1}{M}\sum_{m=1}^{M}P_{i,m},
  \label{eq:12}
\end{eqnarray}
where $S( P_{i})$ represents the prestige score of a scholarly paper, $\langle P_{i,m}\rangle$ represents the average value of higher-order weighted quantum PageRank scores, $M$ represents the iteration number of the algorithm, and $P_{i,m}$ indicates the $m$-th value of higher-order weighted quantum PageRank scores. The concept of the prestige score is inherited from Quantum Google algorithm~\cite{Paparo2014Quantum}, with the importance of a node corresponds to the prestige score of a scholarly paper in our work.

\subsection*{Data description}
Our experiments are conducted on the Physical Review C (PRC) data set, a subset of the American Physical Society (APS) data set (http://publish.aps.org/datasets). PRC consists of 34,443 papers, and each paper includes details of title, author name and affiliation, date of publication, and a list of cited papers. Then, 3,587 papers without citation details from the PRC data set are removed. Overall, 212,421 citations are identified from the data set. Geographic coordinates of over 27,000 institutions are obtained by calling the Geocode function of the Google Maps API.

\subsection*{Data Processing}
To better explore the relationship between citations and geographical distance, we divide geographical distance by adopting statistical analysis technique: grouping analysis and clustering analysis. For grouping analysis, we use multiples of 100 Km distance as threshold values, to determine the group of any two institutions. For instance, if institutes $I_a$ and $I_b$ are 250 Km apart, citations will be considered in the 200--300 Km group. For clustering analysis, we use Density-Based Spatial Clustering of Applications with Noise (DBSCAN)~\cite{Ester1996Density}, which is a spatial clustering algorithm based on density. The number of clusters are determined by two parameters: (1) the furthest distance of any two points belonging to a cluster; and (2) the smallest number of samples in a cluster. We select the DBSCAN for clustering analysis method, mainly because this algorithm has the advantages of fast clustering and efficient processing of noise points and spatial clustering of arbitrary shape.

\subsection*{Geographic distribution of institutions}
\begin{figure*}[!h]
\centering
 \includegraphics[width=0.9\textwidth]{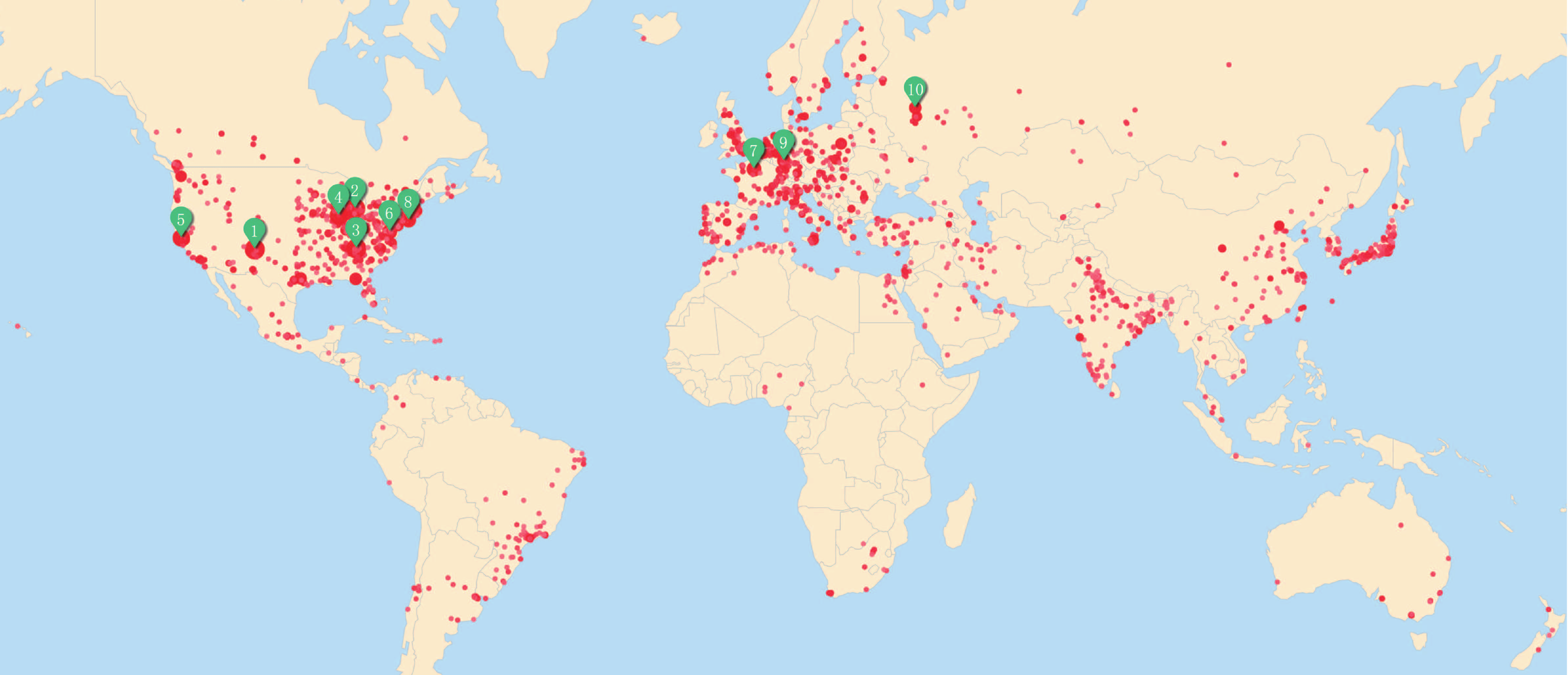}
  \caption{Visualizing the geographical distribution of institutions.}
  \label{Fig.2}
\end{figure*}

Fig~\ref{Fig.2} shows the geographic distribution of institutions with PRC publications. Each red dot represent an institution, with the dot size reflecting the number of publications, and the color represents the number of citations. We observe that research institutions are spread over all continents, with the ones in North America and Europe are more research intensive and attract more citations. The top 10 institutions according to the number of papers published are shown in Table~\ref{tab:1}, and their geographical locations are pointed out by green labels in Fig~\ref{Fig.2}.

\begin{sidewaystable}[hbt!]
\newcommand{\tabincell}[2]{\begin{tabular}{@{}#1@{}}#2\end{tabular}}
\centering
\caption{Top-10 institutions by publication quantities in 1970--2013}
\begin{tabular}{|c|c|c|c|c|c|} \hline
Number&Institution                                                                                                            & Latitude  & Longitude  & Number of papers& Citations\\ \hline
1  &\tabincell{c}{Los Alamos Scientific Laboratory,\\ University of California, \\Los Alamos, New Mexico 87544}            & 35.8800364&-106.3031138&1,723            &14,739\\ \hline
2  &\tabincell{c}{Cyclotron Laboratory and Physics Department,\\ Michigan State University, \\East Lansing, Michigan 48823}& 42.727455 &-84.498557  &1,722            &14,748\\ \hline
3  &\tabincell{c}{Joint Institute for Heavy Ion Research,\\ Oak Ridge, Tennessee 37835, USA}                               & 36.0324131&-84.2316929 &1,703            &13,684\\ \hline
4  &\tabincell{c}{Argonne National Laboratory,\\ Argonne, Illinois 60441, USA}                                             & 41.5945343&-88.0411993 &1,545            &4,128 \\ \hline
5  &\tabincell{c}{Berkeley Geochronology Center,\\ 2455 Ridge Road, Berkeley,\\ California 94709}                          & 37.8759643&-122.2609648&1,414            &10,865\\ \hline
6  &\tabincell{c}{Catholic University of America,\\ Washington, D. C. 20017}                                               & 38.9332429&-76.9945436 &1,133            &9,402 \\ \hline
7  &\tabincell{c}{C.E.N. Saclay, B.P.No. 2,\\ 91190-Gif-sur-Yvette, France}                                                & 48.7082852&2.1620986   &1,132            &8,361 \\ \hline
8  &\tabincell{c}{Departement of Chemistry, \\Brookhaven National Laboratory,\\ Upton, New York, USA}                      & 40.8682379&-72.8791716 &1,118            &9,599 \\ \hline
9  &\tabincell{c}{G.S.I. Darmstadt, 6100 Darmstadt,\\ Germany}                                                             & 49.8685006&8.6493409   &969              &6,414 \\ \hline
10 &\tabincell{c}{Bogoliubov Laboratory of Theoretical Physics,\\ JINR, Dubna, RU-141980 Russia}                           & 56.7417029&37.1911003  &863              &5,965 \\ \hline
%2& 2   &2  &3& 2   &2& 2\\ \hline
%3  &3  &3  &2  &3& 7&  7\\ \hline
%4  &4  &4  &8  &5  &4  &4\\ \hline
%5& 5   &9  &4  &7  &6  &6\\ \hline
%6&6    &5  &5  &4  &3& 3\\ \hline
%7& 7   &8  &7  &6  &12&    15\\ \hline
%8& 8&  13  &6& 9   &5& 5\\ \hline
%9&9&   7   &17 &19&    23& 31\\ \hline
%10&    10& 6   &16 &14 &13&    26\\ \hline
\end{tabular}
\label{tab:1}
\end{sidewaystable}

\section*{Results}
\subsection*{Citation dynamics}
\begin{figure}[!h]
  \centering
  \includegraphics[width=1.0\textwidth]{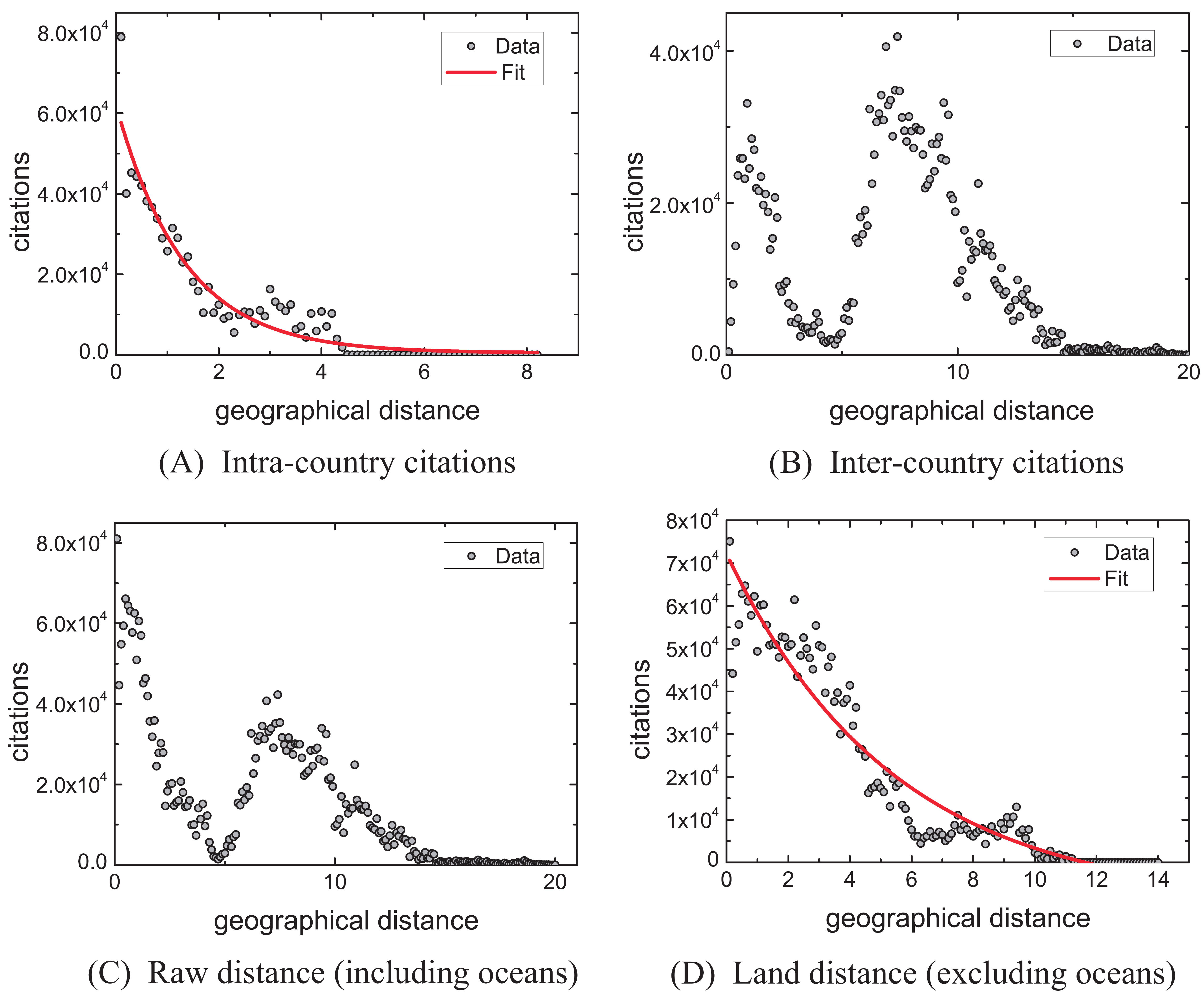}
  \caption{Characterizing the relationship of citations and geographical distance by grouping analysis.}
  \label{Fig.3}
\end{figure}

\begin{figure}[!h]
  \centering
  \includegraphics[width=1.0\textwidth]{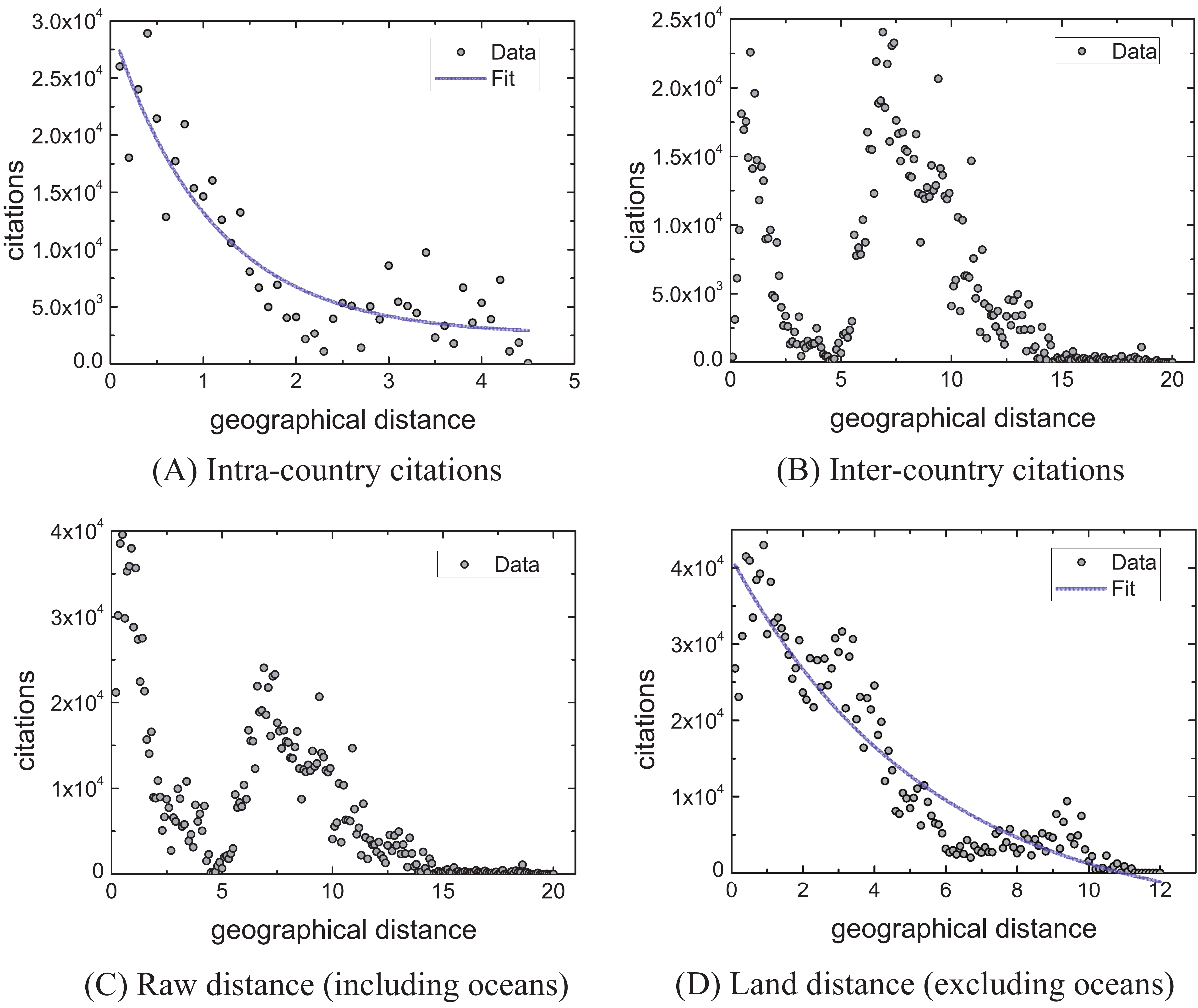}
  \caption{Characterizing the relationship of citations and geographical distance by clustering analysis.}
  \label{Fig.4}
\end{figure}

Fig~\ref{Fig.3} characterizes the change of citations ($C$) (i.e. variation in citation quantity) with geographical distance ($d$) by grouping analysis. In order to characterize the citation trend, we also analyze the relationship between citations and geographical distance by clustering analysis (Fig~\ref{Fig.4}). For both analysis methods, we analyze the citation trend by considering four cases: intra-countries, inter-countries, raw distance (with oceans) and land distance (without oceans). The citation trends of scholarly papers within the countries approximately follows $C(d)$ $\sim$ $y_{0}+A_{1}e^{\frac{-d}{^{t_{1}}}}$ (Fig~\ref{Fig.3}A and Fig~\ref{Fig.4}A). Yet, we find that the citation trends in-between countries (Fig~\ref{Fig.3}B and Fig~\ref{Fig.4}B) are different from the ones within the countries. Citations rapidly decrease when the geographical distances between institutions range from 0 Mm to 5 Mm, then consistently increase and reach the peak at around 7Mm.

The citation trend exhibits a rapid decline from 7 Mm to 20 Mm. Together, Fig~\ref{Fig.3}C and Fig~\ref{Fig.4}C indicate that citations change with actual geographic distance. We find, however, that the changing trend of the citations is similar to one of between countries. This phenomenon drives us to explore the reason behind the peak point in Fig~\ref{Fig.3}B, Fig~\ref{Fig.3}C, Fig~\ref{Fig.4}B, and Fig~\ref{Fig.4}C. As a result, we find that the distance of Atlantic Ocean plays a significant role. The reason is that the Atlantic separates America from Europe, about 75\% affiliations and 67\% citations of papers are from America and Europe, and the citations between America and Europe account for around 68\% of the total citations.
%The large percentage of citations and America and Europe separated by Atlantic are the most critical factors for forming the peak in Fig~\ref{Fig.3}B, Fig~\ref{Fig.3}C, Fig~\ref{Fig.4}B and Fig~\ref{Fig.4}C. Namely,
The uneven geographical distribution of institutions causes such trend. This observation drives us to explore the relationship between citations and geographical distance ignoring the influence of the non-uniform geographical distribution of institutions. To this end, we construct the distance matrix of six continents containing Asia, Europe, Africa, North America, South America, and Oceania through the ranging function of Google Maps. According to Fig~\ref{Fig.3}D and Fig~\ref{Fig.4}D, it is apparent that the change of citations for publications closely relates to the geographical distance, with more citations associate to shorter geographical distance, and vice versa. This has clear implications in quantifying the impact of scholarly papers: if the citations of a paper are from long distance, these citations are more valuable compared to the citations of short distance, and further elaboration can be found in the Discussion Section. Fig~\ref{Fig.3}D and Fig~\ref{Fig.4}D indicate citations appear to follow a similar trend as Fig~\ref{Fig.3}A and Fig~\ref{Fig.4}A.

\begin{figure}[!h]
 \centering
  \includegraphics[width=1.0\textwidth]{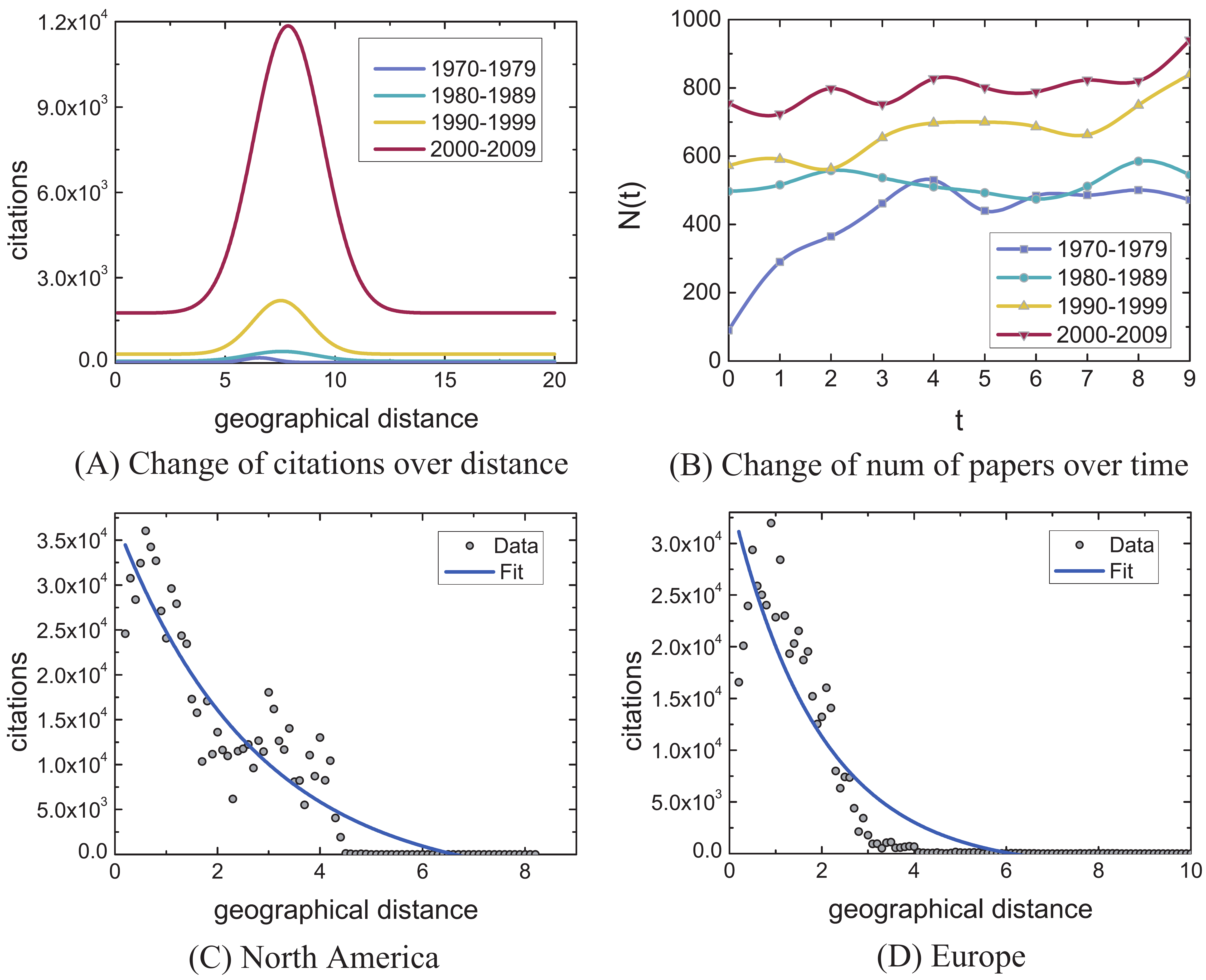}
  \caption{Characterizing citation dynamics.}
  \label{Fig.5}
\end{figure}

In addition, we analyze the citation trend by considering the time factor. To illustrate the difference of citation trend and geographical distance over different periods, comparisons over 4 decades ('70s, '80s, '90s and '00s) are shown in Fig~\ref{Fig.5}. Fig~\ref{Fig.5}A compares the relationship between citations and geographical distance in 4 decades. The trends are shown in Fig~\ref{Fig.3}B, Fig~\ref{Fig.3}C, Fig~\ref{Fig.4}B, and Fig~\ref{Fig.4}C. Fig~\ref{Fig.5}B compares the change of papers over time. Fig~\ref{Fig.5}A indicates that the citation trend approximately follows a Gaussian distribution, ranging from 4 Mm to 12 Mm, and between 2000 to 2009. The trend change is more noticeable compared to other three periods of time: 1970-1979, 1980-1989, 1990-1999. The differences of citation trends (Fig~\ref{Fig.5}A) are consistent with the change of productivity (Fig~\ref{Fig.5}B) in four periods of time. We observe a positive correlation between the number of publications and citations. Fig~\ref{Fig.5}C and Fig~\ref{Fig.5}D show the trends of citation in North America and Europe. These trends indicate that citations are closely related to geographical distance. These results inspire us to evaluate the impact of papers based on the geographical distance (see Methods).

\subsection*{Comparing the impact of papers}
\begin{figure}[!h]
\centering
\includegraphics[width=1.0\textwidth]{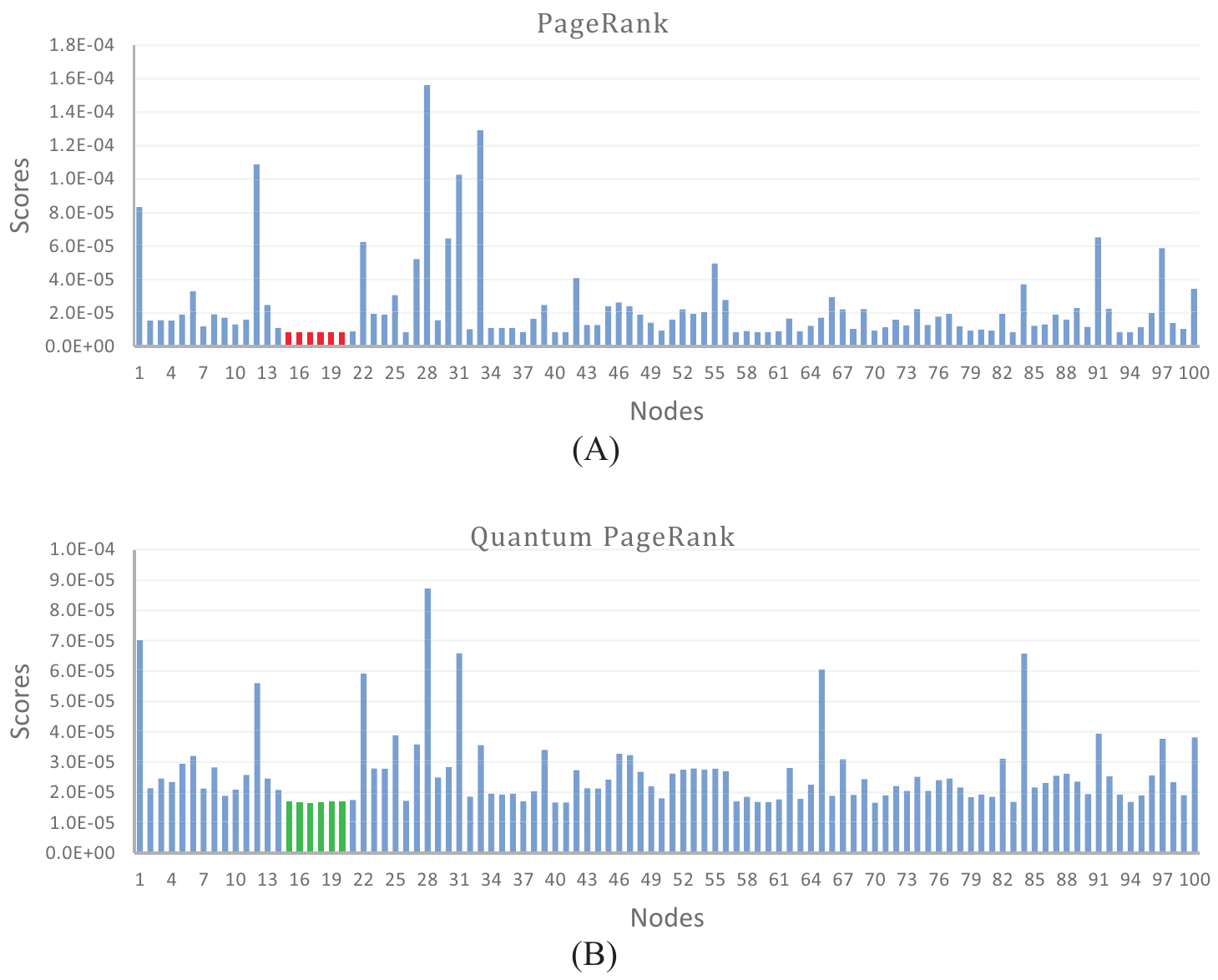}
\caption{Comparing the scores of PageRank and quantum PageRank. Identical scores using PageRank (the red region) can be differentiated using quantum PageRank (the green region).}
\label{Fig.6}
\end{figure}

Based on citation network, we compare the scores of quantum PageRank and PageRank. We observe that many papers share the same PageRank score. The importance of some nodes in citation network cannot be distinguished, which is considered a typical drawback of PageRank. In order to show the difference of scores of quantum PageRank and PageRank, we randomly select 100 scores out of ~27,000 for each algorithm. Fig~\ref{Fig.6} shows the comparative results of quantum PageRank and PageRank for the same nodes in the citation network. According to Fig~\ref{Fig.6}A, we observe that node15 - node21 yield the same scores of PageRank, while their quantum PageRank scores are different (see Fig~\ref{Fig.6}B). Fig~\ref{Fig.6} indicates that quantum PageRank can better reveal the hierarchy of levels in the hierarchical networks.

\begin{figure}[!h]
\centering
\includegraphics[width=1.0\textwidth]{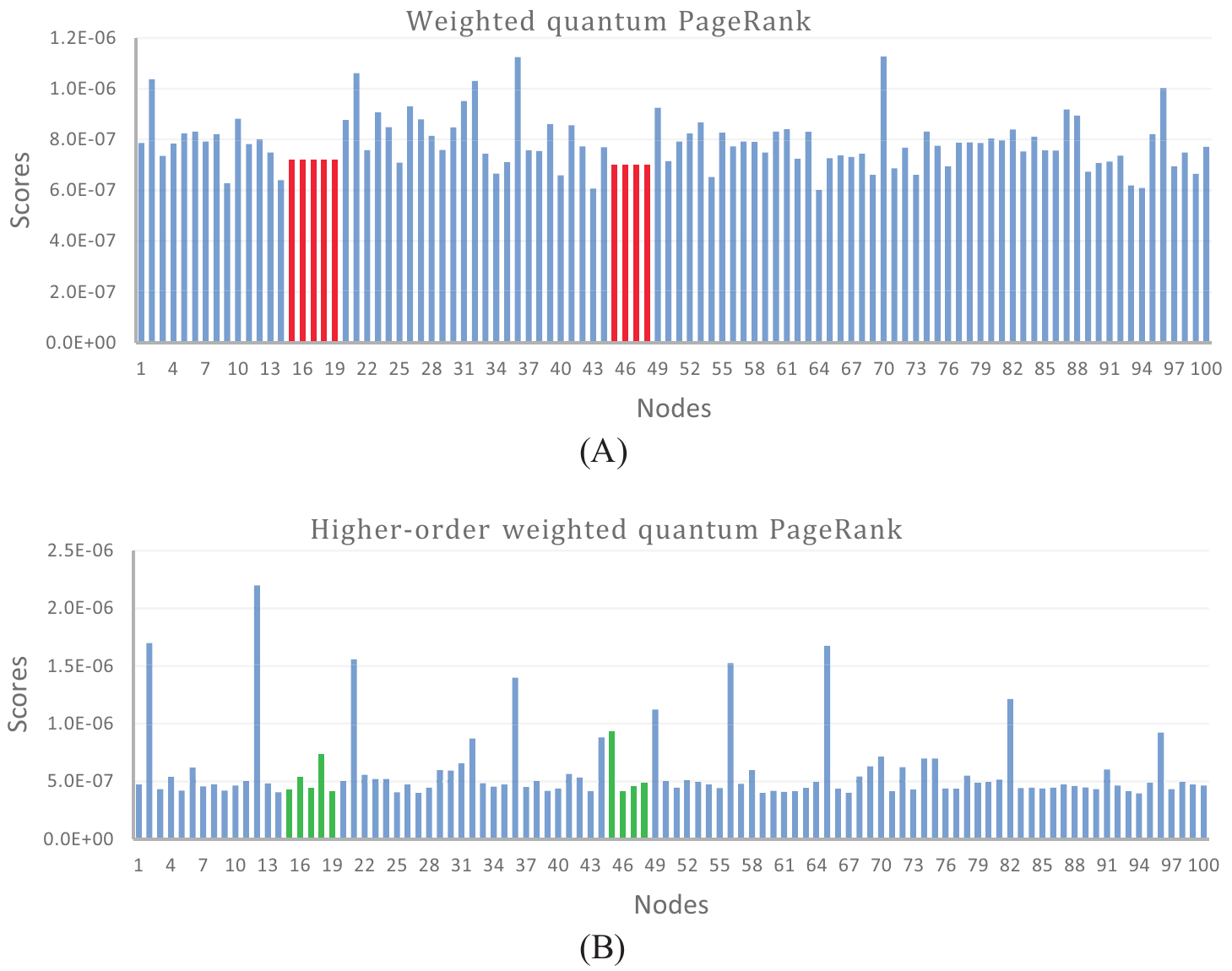}
\caption{Comparing the scores of weighted quantum PageRank and higher-order weighted quantum PageRank. Identical scores using weighted quantum PageRank (the red regions) can be differentiated using higher-order weighted quantum PageRank (the green regions).}
\label{Fig.7}
\end{figure}

In order to explore the performance of higher-order weighted quantum PageRank, we compare the scores of higher-order weighted quantum PageRank and weighted quantum PageRank. We find that higher-order weighted quantum PageRank algorithm can capture different scores when weighted PageRank algorithm shows the same scores, as shown in Fig~\ref{Fig.7}. Fig~\ref{Fig.7}A shows the scores of weighted quantum PageRank of 100 random selected nodes in citation network, and Fig~\ref{Fig.7}B shows the scores of higher-order weighted quantum PageRank of the same nodes. According to this Figure, we observe that scores of node15 - node19 are the same in the weighted quantum PageRank, while their scores are different in the higher-order quantum PageRank. The comparison between the two algorithms indicates that considering higher-order dependencies in citation network can better identify the impact of papers.

\begin{figure}[!h]
\centering
\includegraphics[width=1.0\textwidth]{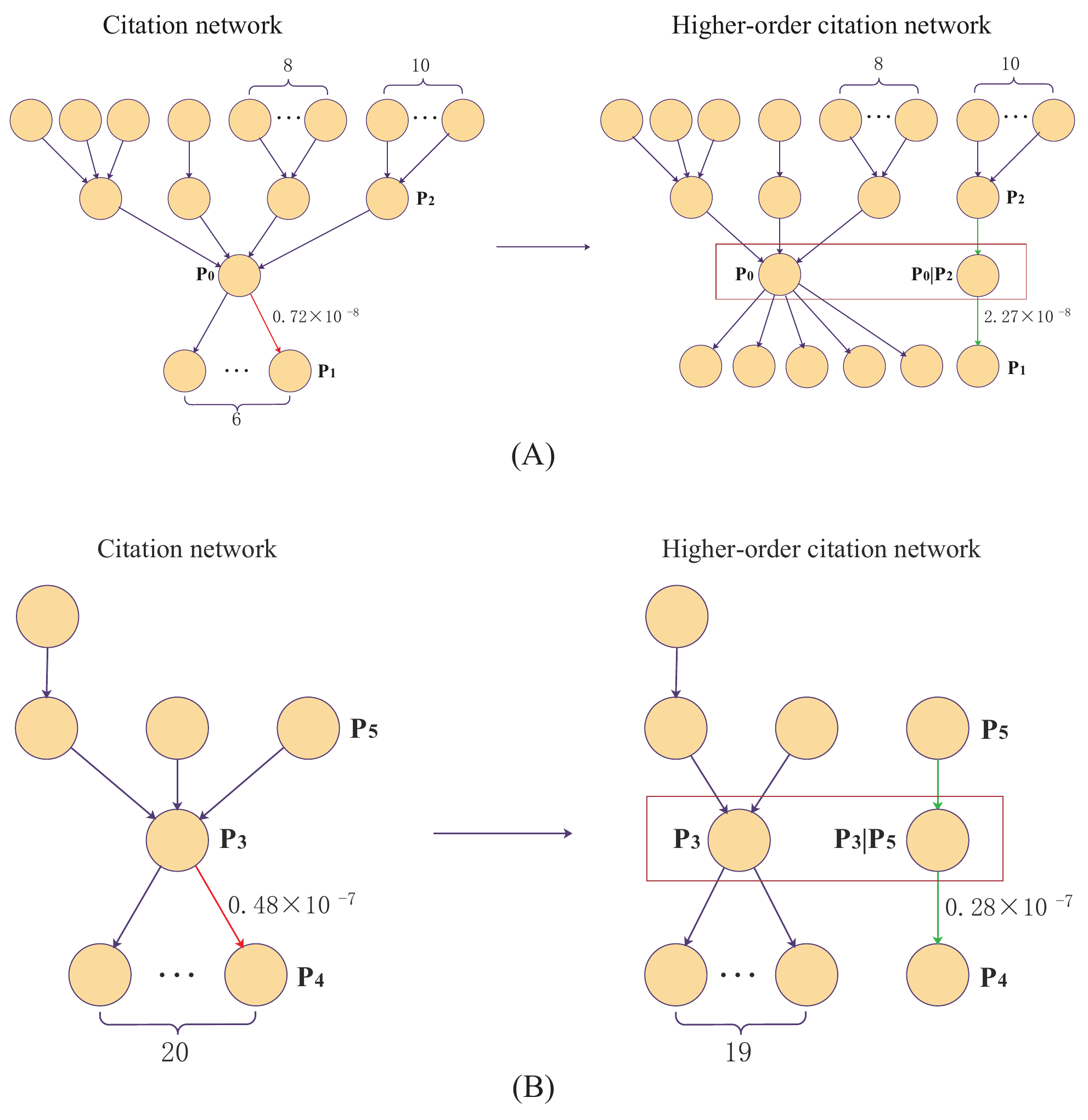}
\caption{Comparing self-citation weights in two different citation networks.}
\label{Fig.8}
\end{figure}

Fig~\ref{Fig.8} illustrates the effect of higher-order citation networks on quantifying the impact of self-citation. The pathways represent the citation between different papers. The red arrow indicates self-citation. The green arrow indicates the self-citation chain with higher-order dependencies. In Fig~\ref{Fig.8}A, paper $P_{0}$ cites paper $P_{1}$, and the citation belongs to self-citation. Other citation relationships do not include self-citation. For example, paper $P_{2}$ cites paper $P_{0}$, and there is no common author for the two papers. Four papers cites $P_{0}$, and $P_{0}$ cites six papers. $W\left ( P_{0}\rightarrow P_{1} \right )$ represents the weight of paper $P_{0}$ cites paper $P_{1}$. $W\left ( P_{0}\rightarrow P_{1} \right )$ is equal to $0.72\times 10^{-8}$ in the original citation network. However, in the higher-order citation network, $W\left ( P_{0}\rightarrow P_{1} \right )$ is equal to the weight of paper $P_{0}|P_{2}$ citing paper $P_{1}$ ($W\left ( P_{0}|P_{2}\rightarrow P_{1} \right )$), namely $2.27\times 10^{-8}$. The weight in the higher-order network is higher than the weight in the original citation network, indicating that the impact of the self-citation is improved . The citation structure contributes to the enhancement of weight of self-citation. Due to the pre-sequence nodes of paper $P_{0}$ are cited multiple times in the citation network, the weight of paper $P_{0}|P_{2}$ citing paper $P_{1}$ is improved in the higher-order citation network. In Fig~\ref{Fig.8}B, paper $P_{3}$ cites paper $P_{4}$, and the citation belongs to self-citation. There is no self-citation in other citation relationships. $W\left ( P_{3}\rightarrow P_{4} \right )$ represents the weight of paper $P_{3}$ cites paper $P_{4}$ in the original citation network. $W\left ( P_{3}|P_{5}\rightarrow P_{4} \right )$ represents the weight of paper $P_{3}$ cites paper $P_{4}$ in the higher-order citation network. We observe that the $W\left ( P_{3}|P_{5}\rightarrow P_{4} \right )$ in the higher-order citation network is lower than $W\left ( P_{3}\rightarrow P_{4} \right )$ in the original citation network. The reason is that pre-sequence nodes of paper $P_{3}$ are only cited by a paper, and paper $P_{5}$ is a root node in the higher-order citation chain. The citation structure determines the weight change in the higher-order citation network.
\section*{Discussion}
\subsection*{Geographical distance}
An interesting finding is that citation pattern is closely related to the geographical distributions of institutions, discounting the separation by oceans. The shorter the actual geographic distance between citing and cited institutions, the more citations. We weight the citation between institutions by ignoring the ocean separating them. Rare citations are considered more valuable: ``less is more.'' Intuitively, long distance presents a barrier for disseminating research finding and socializing other researchers in person. Although publishing over the Internet has become a popular alternative, it is a challenge to promote among massive information made available on the Web. In addition to the Web presence, additional publicity through conferences, seminars, and workshops help making the work well-known. With the increased cost and effort for frequent travel to far-away destinations, citations made by geographically far-away researchers are considered more valuable. At the same time, long-distance citations include less manipulated promotion, thus better reflects the true impact of a paper.

It should be noted that the finding does not conflict to, and can be applied as a weighted factor on-top of, other ``reputation'' metrics such as citations from a paper written by a leading institute or published in a prestigious journal. Investigation of the weighted citation would be a different topic, and to combine it with the geographical distance analysis is beyond the scope of this paper.

\subsection*{Higher-order dependencies}
In this paper, we propose a quantitative approach for evaluating the impact of scholarly papers via a higher-order citation networks.
Evaluating the impact of papers in higher-order citation networks can more objectively reflect the true influence of scholarly papers. Meanwhile, the higher-order dependencies can weaken the effect of manipulated citation activities.
For example, when researchers manipulate citations to boost the impact of their papers, they usually deliberately cite the new published papers by themselves or their friends. The manipulation activities can influence the true citation networks, and generate more influence to the first-order citation networks.

The higher-order dependencies are more likely to happen for the denser nodes and root nodes in citation networks. We exclude sparse nodes (citation chains with appearing less than 50 times in all the citation chains) in the citation networks to find the higher-order dependencies. The ignored nodes in citation networks are regarded as the zero-order dependencies, and such nodes are a large proportion in citation networks. In fact, the number of citation relationships is based on the statistical citation chains, which is generated by using the random walk method. Therefore, for a certain pair of citation, we find that the number of cited papers of precedence nodes and the number of citations of the succeeding nodes determine the number of occurrences of the pair of nodes in all the citation chains. Based on this finding, we roughly estimate that the probability of such nodes getting more citations is low if the higher-order dependencies of the nodes appear in the citation chains less than 50 times. Given a paper, we trace its citation path, and we generate a citation tree according the citation relationships.  For the root node in the citation tree, the total number of the root nodes appearing in all the citation chains is only related to the post-sequence nodes. For the leaf node in the citation tree, the total number of the leaf nodes appearing in all the citation chains is only related to the pre-sequence nodes. The finding mentioned above can be extended to all the networks, in which researches can find the corresponding higher-order dependencies to better rank the nodes. The general pattern is that the number of in-degree and the number of out-degree of a node determine the number of occurrences of the node in all the communication paths in certain network. Furthermore, the general pattern can be used to evaluate the importance of nodes in different networks.

\subsection*{Ranking algorithm analysis}
Due to the scores of PageRank more depending on the damping parameter $\alpha$, the scores of PageRank look more arbitrary. Compared to PageRank, the scores of quantum PageRank are less dependent on the parameter $\alpha$, indicating quantum PageRank is more robust compared to PageRank in term of the variation of damping parameter $\alpha$~\cite{Paparo2014Quantum}. We find that more citations are associated to shorter geographical distances. To weaken the impact of cited papers from citing papers with short distances, and strengthen the impact of scholarly papers from citations with long distances, we weight the citation networks by an inverse function of the geographical distance between institutions. Based on the finding that citations are closely related to geographical distance, we construct the higher-order weighted quantum PageRank algorithm for objectively quantifying the impact of scholarly papers. In the hierarchical networks, quantum PageRank can better distinguish the impact of nodes compared to PageRank, as shown in Fig~\ref{Fig.6}. Higher-order weighted quantum PageRank can capture deeper structured information, and better distinguish the impact of nodes compared to weighted quantum PageRank, as shown in Fig~\ref{Fig.7}.
%\section*{Conclusion}
%
%CO\textsubscript{2} Maecenas convallis mauris sit amet sem ultrices gravida. Etiam eget sapien nibh. Sed ac ipsum eget enim egestas ullamcorper nec euismod ligula. Curabitur fringilla pulvinar lectus consectetur pellentesque. Quisque augue sem, tincidunt sit amet feugiat eget, ullamcorper sed velit.
%
%Sed non aliquet felis. Lorem ipsum dolor sit amet, consectetur adipiscing elit. Mauris commodo justo ac dui pretium imperdiet. Sed suscipit iaculis mi at feugiat. Ut neque ipsum, luctus id lacus ut, laoreet scelerisque urna. Phasellus venenatis, tortor nec vestibulum mattis, massa tortor interdum felis, nec pellentesque metus tortor nec nisl. Ut ornare mauris tellus, vel dapibus arcu suscipit sed. Nam condimentum sem eget mollis euismod. Nullam dui urna, gravida venenatis dui et, tincidunt sodales ex. Nunc est dui, sodales sed mauris nec, auctor sagittis leo. Aliquam tincidunt, ex in facilisis elementum, libero lectus luctus est, non vulputate nisl augue at dolor. For more information, see \nameref{S1_Appendix}.
\section*{Supporting information}
%
%% Include only the SI item label in the paragraph heading. Use the \nameref{label} command to cite SI items in the text.
%\paragraph*{S1 Fig.}
%\label{S1_Fig}
%{\bf Bold the title sentence.} Add descriptive text after the title of the item (optional).
%
%\paragraph*{S2 Fig.}
%\label{S2_Fig}
%{\bf Lorem ipsum.} Maecenas convallis mauris sit amet sem ultrices gravida. Etiam eget sapien nibh. Sed ac ipsum eget enim egestas ullamcorper nec euismod ligula. Curabitur fringilla pulvinar lectus consectetur pellentesque.
%
\paragraph*{S1 Data Source.}
\label{S1_Data} {\bf Data source used in this paper.}

%{\bf Lorem ipsum.}  Maecenas convallis mauris sit amet sem ultrices gravida. Etiam eget sapien nibh. Sed ac ipsum eget enim egestas ullamcorper nec euismod ligula. Curabitur fringilla pulvinar lectus consectetur pellentesque.
%
%\paragraph*{S1 Video.}
%\label{S1_Video}
%{\bf Lorem ipsum.}  Maecenas convallis mauris sit amet sem ultrices gravida. Etiam eget sapien nibh. Sed ac ipsum eget enim egestas ullamcorper nec euismod ligula. Curabitur fringilla pulvinar lectus consectetur pellentesque.
%
%\paragraph*{S1 Appendix.}
%\label{S1_Appendix}
%{\bf Lorem ipsum.} Maecenas convallis mauris sit amet sem ultrices gravida. Etiam eget sapien nibh. Sed ac ipsum eget enim egestas ullamcorper nec euismod ligula. Curabitur fringilla pulvinar lectus consectetur pellentesque.
%
%\paragraph*{S1 Table.}
%\label{S1_Table}
%{\bf Lorem ipsum.} Maecenas convallis mauris sit amet sem ultrices gravida. Etiam eget sapien nibh. Sed ac ipsum eget enim egestas ullamcorper nec euismod ligula. Curabitur fringilla pulvinar lectus consectetur pellentesque.

\section*{Acknowledgments}
The authors extend their appreciation to the International Scientific Partnership Program ISPP at King Saud University for funding this research work through ISPP\#0078. The funders had no role in study design, data collection and analysis, decision to publish, or preparation of the manuscript.

\end{document}